\documentclass[url,12pt]{article}
\usepackage[toc,page]{appendix}
\pdfoutput=1
\usepackage{graphicx,pdflscape,amsmath}
\usepackage{textcomp}
\usepackage{alltt,dsfont,amsmath,amssymb,bm}
\usepackage{color}
\usepackage{url}
\usepackage{pdfpages}

\newcommand{\mma}{ {\em Mathematica }}
\newcommand{\dirac}{{\bf \color{red} DiracQ} }

\newcommand{\beq}{\begin{equation}}
\newcommand{\eeq}{\end{equation}}
\newcommand{\barray}{\begin{eqnarray}}
\newcommand{\earray}{\end{eqnarray}}

\begin{document}
\title{{\bf {\huge \dirac} \\ {\Large A Quantum Many-Body Physics Package }  }
\footnote{
 \textcircled{c} J. G. Wright and B. S. Shastry (2013). \newline
 Distributed under  the GNU General Public License, version 3. \newline
 {{\bf  http://www.gnu.org/licenses/gpl-3.0-standalone.html}}
  } }
\author{{\bf John G. Wright and B. Sriram Shastry}\\ \\
{\small Physics Department} \\
{\small University of California  Santa Cruz, CA 95064}}
\date{{\small January 14, 2013} \\ \vspace{.5in}
Website of \dirac    \\ \vspace{.25in}
\url{http://DiracQ.org} \\
{\small Or}\\
\url{http://physics.ucsc.edu/~sriram/DiracQ}
}
\maketitle
\newpage
 {\bf \large $\blacklozenge$   1. Introduction:}
\vspace{.25in}

We present a software package \dirac, for use in quantum many-body Physics. It is designed for helping with typical algebraic manipulations that arise in quantum Condensed Matter Physics and Nuclear Physics problems, and also in some subareas of Chemistry. DiracQ is invoked within a \mma session, and extends the symbolic capabilities of \mma by building in standard commutation and anticommutation rules for several objects relevant in many-body Physics. It enables the user to carry out computations such as evaluating the commutators of arbitrary combinations of spin, Bose and Fermi operators defined on a discrete lattice, or the position and momentum operators in the continuum. Some examples from popular systems, such as the Hubbard model, are provided to illustrate the capabilities of the package.

A word about the underlying philosophy of this program is appropriate here.  Physicists approach  calculations in quantum theory  with  a certain  informality that
 contrasts with  the approach  of  mathematicians, who  usually insist on a somewhat  more rigorous (notational)
 framework. Some formality is also   present   in the method of programming  a computation with  older computer languages  such as Fortran.  The physicist's informality refers specifically to a deferment of definitions and properties of objects, until one actually needs them. 
This enables one to write down and manipulate and to compound expressions, often  into simple and  useful final  results.  In our view the  physicist's approach owes much to the notation introduced by  Dirac  in 1927. In the classic paper  ``The Physical Interpretation of Quantum Dynamics'', P. A. M. Dirac  Proc. Roy. Soc.{\bf  A113},621(1927), we find the first mention of  c-numbers and q-numbers as follows:

{\em \ldots one cannot suppose the dynamical variables to be ordinary numbers (c-numbers), but may call them numbers of a special type (q-numbers).} 

 This inspired  notation helps   us  to treat the commuting parts of the expression with standard care (due  their c-number nature), while reserving extra  care  for handling  the q-number parts.

 In \dirac,  we  initially declare a set of symbols to be operators, i.e. q-numbers.  The properties of standard operators such as Bose, Fermi, position-momentum, angular momentum etc are  already programmed, and  if required, a user could define more exotic operators with specific properties. Once this is done,    any expression is split into its c-number  and q-number parts as the first and fundamental operation. Complicated input  expressions may be written with a fairly informal syntax,    with mixtures of c-numbers  and q-numbers, and as  the  sums of such expressions. \dirac     handles them  by first separating them into their c-number  and q-number parts.   Standard operations on   operators, such as adding, multiplying or  commuting  them  is  then straightforward, the c-number parts are spectators for most part while the q-numbers  are combined using the given rules.  Finally the expressions are written back in standard  input like notation. The package \dirac consists of a set of mutually dependent  functions to perform most of the operations. These functions are most often named using  the last letter as  ``Q''; 
 for example we denote a  function (described  later) as  SimplifyQ[a], thus  distinguishing  it from the function Simplify[a] of \mma.

 This  version  of \dirac  uses commands that call  upon the symbolic capabilities of {\em Mathematica},
 and exploits  its ability to deal with non commuting symbols.  It might be  possible to couple the commands of \dirac with an OSS program such as Sage, in view of the relatively small fraction of  commands of \mma that are actually used here. The present version of \dirac was developed and  tested on Version 8  of {\em Mathematica}, and requires the palettes feature of this version in order to declare the standard non commuting symbols. A version avoiding the palettes is planned for the future.

\vspace{.25in}
{\bf \large $\blacklozenge$ 2. Contents:}
\vspace{.15in}

The package distribution contains the following files: ($X$ is the version number)
\begin{enumerate}
\item{\bf Introduction.pdf }

This article.
\item {\bf DiracQ\_VX.m} 

The main package implementing \dirac. 
\item {\bf GettingStarted\_VX.nb}

 A first introduction to loading \dirac and simple examples.
\item{ \bf Glossary\_VX.nb} 

A list of all  commands in \dirac, their description  and a simple example or two of  their usage. It also contains, under the heading {\bf DiracQPalette}, a description of the palette used in this notebook, with its various predefined operators and instructions on how to ``turn them on". 
\item{\bf Tutorial\_VX.nb} 

A more detailed introduction to the commands and  their illustration in simple problems. This notebook has some overlap with the notebook Glossary\_VX.nb, it provides an in depth commentary on the usage of the various functions defined in the package. Simple examples with harmonic oscillator operators are provided as warming up. Examples are provided for all the  functions available in \dirac.
 Users who would like to enlarge the class of operators and define their own set, would find a helpful example treated, where the defining relations of the Virasoro  algebra are added, and a few simple computations with their generators carried out. 
\item{\bf Examplebook\_VX.nb}

A set of calculations  using of \dirac in  problems of various level of difficulty. 
Harmonic oscillator commutations provide a simple introduction to the package, followed by a demonstration of
the conservation of the Runge-Lenz-Laplace vector in the Hydrogen atom.  The Bra and Ket objects of \dirac are introduced. 
 Fermionic operators are illustrated in the context of the Hubbard model, where the commutators of  combinations of Fermi  operators with the Hamiltonian are evaluated,  leading to exact expressions for  the exact second and third  moments of the electron  spectral function. Applications to spin half particles and Pauli matrices are given using R. J. Baxter's celebrated   proof (1971) of the integrability of the 8-vertex model.
   The relations between the  Boltzmann weights required for   satisfying  the   star triangle  relations are recovered  using \dirac.  Also provided is a more complex example based on the 1-d Hubbard model. Its  S matrix   ( Shastry 1986) is known, which  can further   be used to construct an inhomogeneous generalization of the Hubbard model. This construction is dependent upon the S matrix  satisfying  a more complex relation that is very hard to check analytically (due to the large number of terms involved). It  can be verified  easily using \dirac, providing  a nontrivial  demonstration.  Another example illustrates the higher conservation laws of the 1-d Hubbard model that are non quadratic in the Fermi operators. 

\dirac can also be used in the initial stages  of  numerical calculations on finite systems, since it can   generate the numerical  Hamiltonian matrix on a given cluster. This is illustrated in a $4 $  site cluster with nearest and next nearest hopping bonds in the sector of 2 up and 2 down particles. This example also  illustrates the strategies  for generation of states using the Bra and Ket objects of \dirac. 

\end{enumerate}

\vspace{.5in}
{\bf \large $\blacklozenge$ 3. Contacts:}
\vspace{.25in}

The authors can be contacted with comments and suggestions as well as further user generated  examples of \dirac  by many channels.  The preferred one is the email address  for this purpose :
\begin{center}
{ \bf DiracQ@gmail.com} 
\end{center}

\vspace{.5in}
{\bf \large $\blacklozenge$ 4. Suggested  Acknowledgement of  \dirac }
\vspace{.25in}

If \dirac is useful in obtaining results leading to publication, a citation would be appropriate and appreciated.
 
\vspace{.25in}
{\bf Suggested Citation} :

 This work contains results obtained by using  {\em DiracQ: A Quantum Many-Body Physics Package },   J. G. Wright and B. S. Shastry, arXiv:1301.  (2013). 

\vspace{.5in}
 {\bf \large $\blacklozenge$   5. Acknowledgement:}
\vspace{.25in}

 This project was supported  in its preliminary stage   by DOE under Grant No. FG02-06ER46319.

\newpage


\section*{Supplementary material}

# DiracQ Example Book

---

## A Compilation of Problems in Quantum Theory Performed Using the DiracQ Package

The following problems demonstrate how the DiracQ package can be used to solve non-trivial problems. Throughout this notebook ensure that you have the relevant operators activated on the DiracQ palette before evaluating any input.

```
SetDirectory[NotebookDirectory[]];
Get["DiracQ_V0.m"];
```

■ **(I) Canonical Pairs**

The package contains canonical position and momentum operators. Their use is demonstrated below for some simple examples.

```
? p
```

> p is the canonical momentum operator. This operator can be called with one argument, taken to be
>     site index, or two arguments. The second argument will be taken to be coordinate direction. Also
>     included is the 3 dimensional canonical momentum vector, represented by OverVector[p], or $\vec{p}$.

```
? q
```

> q is the canonical position operator. This operator can be called with one argument, taken to be
>     site index, or two arguments. The second argument will be taken to be coordinate direction.
>     Also included is the 3 dimensional canonical position vector, represented by OverVector[q], or $\vec{q}$.

The physical system can be in arbitrary dimensions and may refer to an arbotrary number of particles. The canonical pairs are endowed with indices:  either one or two indices can be used. The first index is taken to be site label. The second index is taken to be the component. If only one index is given, a 1-dimensional system is assumed. This is demonstrated below. [Note: *unless you choose to declare q and p as active operators in the DiracQ Palette by ticking the boxes in the Palette, (or more aggressively turn all symbols on), the following commutators will turn out to vanish. If that happens, go back and turn them  "on".* ]

```
Commutator[p[j], q[i]]
```

$-i \hbar \delta[i, j]$

```
Commutator[p[j, α], q[i, β]]
```

$-i \hbar \delta[i, j] \delta[\alpha, \beta]$

```
Commutator[p[j, y], q[i, y]]
```

$-i \hbar \delta[i, j]$

```
Commutator[p, q]
```

0

Comment: In the last output, notice that p and q without any indices are treated as c-numbers.  This can be a slight nuisance, but one  resolution is to treat a set of  harmonic oscillators indexed by a "i".

```
Clear[i, j]
```

$$H = \frac{p[i]^2}{2 m} + \frac{m\omega^2}{2} q[i]^2;$$

```
Commutator[H, p[i]]
```

$\mathtt{i}\ \left(\mathtt{m}\,\omega^2\ \hbar\right)\ **\ \mathtt{q[i]}$

Next we construct creation and destruction operators from the canonical pairs.

```
a[i_] := √ (m ω / 2 ℏ)  (q[i] + i p[i] / m ω);

a†[i_] := √ (m ω / 2 ℏ)  (q[i] - i p[i] / m ω);

Commutator[a[i], a†[i]]
```

1

```
n[i_] := a†[i] ** a[i]
```

Here we show that [n[i], a[i]] = -a[i] and similarly [n[i],a†[i]]=a†[i]. SimplifyQ puts the expressions in a standard form where the identity of two expressions becomes clear.

```
SimplifyQ[a[i]]
```

$$\frac{\sqrt{\frac{\mathtt{m}\,\omega}{\hbar}}\ **\ \mathtt{q[i]}}{\sqrt{2}}\ +\ \frac{\mathtt{i}\ \frac{1}{\sqrt{\frac{\mathtt{m}\,\omega}{\hbar}}\ \hbar}\ **\ \mathtt{p[i]}}{\sqrt{2}}$$

```
SimplifyQ[Commutator[n[i], a[i]] + a[i]]
```

0

```
SimplifyQ[Commutator[n[i], a†[i]] - a†[i] ]
```

0

■ **(II) Hydrogen atom and Runge Lenz Laplace vectors**

Here we define the Hamiltonian of the Hydrogen atom. We demonstrate the conservation of the Runge-Lenz-Laplace vector by using the package.

We use the three dimensional vector notation for the canonical pair. For example

```
q⃗[i]
```

{q[i, x], q[i, y], q[i, z]}

```
q⃗[i].q⃗[i]
```

$\mathtt{q[i, x]}^2 + \mathtt{q[i, y]}^2 + \mathtt{q[i, z]}^2$

```
Clear[H];

H = p⃗[i].p⃗[i] / 2 m - Z e² / √(q⃗[i].q⃗[i]) ;

? NCcross
```

Non Commutative cross product of two 3dimensional vectors retaining the order of the operators.

```
L⃗[i_] = NCcross[q⃗[i], p⃗[i]]
```

{q[i, y] ** p[i, z] - q[i, z] ** p[i, y],
 -q[i, x] ** p[i, z] + q[i, z] ** p[i, x], q[i, x] ** p[i, y] - q[i, y] ** p[i, x]}

```
Commutator[L⃗[i][[1]], H]
```

0

```
Commutator[L⃗[i][[2]], H]
```

0

```
SimplifyQ[Commutator[L⃗[i][[3]], H]]
```

0

The answer is so simple that it looks as though we have not performed any operation whatsoever. However, if we leave basic commutators unevaluated we see that indeed the computation was not entirely trivial (feel free to try it by setting "Apply Definition" to "False" using the palette and then reevaluate the input). As an example consider the product q[i,y]**p[i,z], it is not conserved and we get a messy answer:

```
Commutator[q[i, y] ** p[i, z], H]
```

$$i \frac{\hbar}{m} ** p[i, y] ** p[i, z] - i \left(e^2 \, Z \, \hbar\right) ** \frac{1}{\left(q[i, x]^2 + q[i, y]^2 + q[i, z]^2\right)^{3/2}} ** q[i, y] ** q[i, z]$$

We next define the Runge Lenz vector $\vec{A}[i]$ with a formal definition:

$$\vec{A}[i] = \vec{L}[i] \times \vec{p}[i] + Z \, e^2 \, m \frac{\vec{q}[i]}{\sqrt{\vec{q}[i].\vec{q}[i]}}$$

Using the NCcross function and symmetrizing in the two variables L and p, we work with

$$\vec{A}[i] = \frac{1}{2} \, \text{NCcross}[\vec{L}[i], \, \vec{p}[i]] - \frac{1}{2} \, \text{NCcross}[\vec{p}[i], \vec{L}[i]] + Z \, e^2 \, m \frac{\vec{q}[i]}{\sqrt{\vec{q}[i].\vec{q}[i]}};$$

We check a component of A to make sure it makes sense.

```
A⃗[i][[3]]
```

$$\frac{1}{2} \, (-p[i, x] ** (-q[i, x] ** p[i, z]) + p[i, y] ** (-q[i, z] ** p[i, y]) -$$

$$\quad p[i, x] ** q[i, z] ** p[i, x] + p[i, y] ** q[i, y] ** p[i, z]) +$$

$$\frac{1}{2} \, (-(-q[i, x] ** p[i, z]) ** p[i, x] + (-q[i, z] ** p[i, y]) ** p[i, y] +$$

$$\quad q[i, y] ** p[i, z] ** p[i, y] - q[i, z] ** p[i, x] ** p[i, x]) + \frac{e^2 \, m \, Z \, q[i, z]}{\sqrt{q[i, x]^2 + q[i, y]^2 + q[i, z]^2}}$$

**term1 = Commutator$\left[\hat{A}[i][[3]], H\right]$**

$\frac{1}{2}$ i $\left(e^2\, Z\, \hbar\right)$ ** p[i, z] ** $\frac{1}{\sqrt{q[i, x]^2 + q[i, y]^2 + q[i, z]^2}}$ +

$\frac{1}{2}$ i $\left(e^2\, Z\, \hbar\right)$ ** $\frac{1}{\sqrt{q[i, x]^2 + q[i, y]^2 + q[i, z]^2}}$ ** p[i, z] +

$\left(e^2\, Z\, \hbar^2\right)$ ** $\frac{1}{\left(q[i, x]^2 + q[i, y]^2 + q[i, z]^2\right)^{3/2}}$ ** q[i, z] $-$

$\frac{1}{2}$ i $\left(e^2\, Z\, \hbar\right)$ ** p[i, x] ** $\frac{1}{\left(q[i, x]^2 + q[i, y]^2 + q[i, z]^2\right)^{3/2}}$ ** q[i, x] ** q[i, z] $-$

$\frac{1}{2}$ i $\left(e^2\, Z\, \hbar\right)$ ** p[i, y] ** $\frac{1}{\left(q[i, x]^2 + q[i, y]^2 + q[i, z]^2\right)^{3/2}}$ ** q[i, y] ** q[i, z] $-$

$\frac{1}{2}$ i $\left(e^2\, Z\, \hbar\right)$ ** p[i, z] ** $\frac{1}{\left(q[i, x]^2 + q[i, y]^2 + q[i, z]^2\right)^{3/2}}$ ** q[i, z] ** q[i, z] $+$

$\frac{1}{2}$ i $\left(e^2\, Z\, \hbar\right)$ ** $\frac{1}{\left(q[i, x]^2 + q[i, y]^2 + q[i, z]^2\right)^{3/2}}$ ** p[i, x] ** q[i, x] ** q[i, z] $+$

$\frac{1}{2}$ i $\left(e^2\, Z\, \hbar\right)$ ** $\frac{1}{\left(q[i, x]^2 + q[i, y]^2 + q[i, z]^2\right)^{3/2}}$ ** p[i, y] ** q[i, y] ** q[i, z] $-$

i $\left(e^2\, Z\, \hbar\right)$ ** $\frac{1}{\left(q[i, x]^2 + q[i, y]^2 + q[i, z]^2\right)^{3/2}}$ ** p[i, z] ** q[i, x] ** q[i, x] $-$

i $\left(e^2\, Z\, \hbar\right)$ ** $\frac{1}{\left(q[i, x]^2 + q[i, y]^2 + q[i, z]^2\right)^{3/2}}$ ** p[i, z] ** q[i, y] ** q[i, y] $-$

$\frac{1}{2}$ i $\left(e^2\, Z\, \hbar\right)$ ** $\frac{1}{\left(q[i, x]^2 + q[i, y]^2 + q[i, z]^2\right)^{3/2}}$ ** p[i, z] ** q[i, z] ** q[i, z]

With this expression, SimplifyQ does not produce a vanishing result since the expression on right locates p[i,z] differently in various terms and one needs to push this term to one side in ALL terms. Therefore at this stage it is better to push  all the momentum type variables to the right extreme ( the left extreme works equally well), so that all operators are ordered in a similar  way. After that  we can set all non commutative products to simple products and simplify.  The function pp and ppp achieve this ordering, and sim does the job of simplifying the expressions after replacing the **'s with simple *'s.

**pp[a\_, m\_] := PushOperatorRight[a, p[i, m]]**

**ppp[a\_] := pp[pp[pp[a, x], y], z]**

**sim[a\_] := Simplify[ppp[a] /. {NonCommutativeMultiply → Times}]**

**sim[term1]**

0

This can be examined more closely by examining the intermediate terms:

```
ppp[term1] /. {NonCommutativeMultiply → Times}
```

$$\frac{3\,e^2\,Z\,\hbar^2\,q[i,\,x]^2\,q[i,\,z]}{2\,\left(q[i,\,x]^2+q[i,\,y]^2+q[i,\,z]^2\right)^{5/2}}+\frac{3\,e^2\,Z\,\hbar^2\,q[i,\,y]^2\,q[i,\,z]}{2\,\left(q[i,\,x]^2+q[i,\,y]^2+q[i,\,z]^2\right)^{5/2}}+$$

$$\frac{3\,e^2\,Z\,\hbar^2\,q[i,\,z]^3}{2\,\left(q[i,\,x]^2+q[i,\,y]^2+q[i,\,z]^2\right)^{5/2}}-\frac{i\,e^2\,Z\,\hbar\,p[i,\,z]\,q[i,\,x]^2}{\left(q[i,\,x]^2+q[i,\,y]^2+q[i,\,z]^2\right)^{3/2}}-$$

$$\frac{i\,e^2\,Z\,\hbar\,p[i,\,z]\,q[i,\,y]^2}{\left(q[i,\,x]^2+q[i,\,y]^2+q[i,\,z]^2\right)^{3/2}}-\frac{3\,e^2\,Z\,\hbar^2\,q[i,\,z]}{2\,\left(q[i,\,x]^2+q[i,\,y]^2+q[i,\,z]^2\right)^{3/2}}-$$

$$\frac{i\,e^2\,Z\,\hbar\,p[i,\,z]\,q[i,\,z]^2}{\left(q[i,\,x]^2+q[i,\,y]^2+q[i,\,z]^2\right)^{3/2}}+\frac{i\,e^2\,Z\,\hbar\,p[i,\,z]}{\sqrt{q[i,\,x]^2+q[i,\,y]^2+q[i,\,z]^2}}$$

```
Simplify[%]
```

0

The same scheme is demonstrated by pushing operators to the left (rather than right) here.

```
ppl[a_, m_] := PushOperatorLeft[a, p[i, m]]

pppl[a_] := ppl[ppl[ppl[a, x], y], z]

siml[a_] := Simplify[pppl[a] /. {NonCommutativeMultiply → Times}]

siml[term1]
```

0

■ **(III) Fermions at work**

● **For this Section Ensure that Fermionic Operators as well as Bra and Ket Vectors are Activated.**

Here we use 1 to represent spin up and - 1 to represent spin down. We now perform some test operations in the context of three problems using the Hubbard model Hamiltonian, given below.

```
Clear[i, j, n, H];
H = - ∑ ∑ ∑ t[i, j] f†[i, σ] ** f[j, σ] + U ∑ n_f[i, 1] ** n_f[i, -1];
         i  j  σ                              i
```

■ **Problem 1 : $[f_i^\sigma, H]$**

These examples require the user to input expressions carefully. For example, notice that we have a sum over $\sigma$ in the first term of H above. We therefore cannot use an operator that has a spin specified by $\sigma$, because the package does not differentiate that one sigma is inside the sum and the other is outside. It will attempt to sum over both $\sigma$'s. This is not so much a restriction as it is a warning against sloppy notation that could easily be understood by a human but will be misinterpreted by the package. The same restriction applies to site indices, such as i and j in this example.

The example below demonstrates incorrect usage of indices; we are taking the commutator of a fermi operator with a site index that is identical to the indice of summation in the Hamiltonian.

```
Commutator[f[i, σ], H]
```

$$\sum_i \left(U\,\delta[1,\,\sigma]\right) ** n_f[i,\,-1] ** f[i,\,1] -$$

$$\sum_i \left(U\,\delta[-1,\,\sigma]\right) ** f†[i,\,1] ** f[i,\,-1] ** f[i,\,1] - \sum_i \sum_j \sum_\sigma t[i,\,j] ** f[j,\,\sigma]$$

The above answer is **not** correct.

Trying same the problem again using proper notation where the external indices are different from the ones used in H,

```
Commutator[f[k, σ1], H]
```

$(U \delta[1, \sigma 1]) ** n_f[k, -1] ** f[k, 1] -$
$\quad (U \delta[-1, \sigma 1]) ** f\dagger[k, 1] ** f[k, -1] ** f[k, 1] - \sum_j t[k, j] ** f[j, \sigma 1]$

we get the correct answer. We can simplify the answer further by specifying a value for $\sigma 1$.

```
Commutator[f[k, σ1], H] /. σ1 → 1
```

$U ** n_f[k, -1] ** f[k, 1] - \sum_j t[k, j] ** f[j, 1]$

The best way of performing this calculation is just to start with the spin specified. In general this method is preferred because the package is then able to organize the operators using the value of the spins. It may therefore be necessary to reorder the output again if you specify the value of a spin after calculation.

```
Commutator[f[k, 1], H]
```

$U ** n_f[k, -1] ** f[k, 1] - \sum_j t[k, j] ** f[j, 1]$

```
Commutator[f[k, -1], H]
```

$-U ** f\dagger[k, 1] ** f[k, -1] ** f[k, 1] - \sum_j t[k, j] ** f[j, -1]$

■ **Problem 2 : $[f_i^\sigma n_i^{-\sigma}, H]$**

In this example we see that the answer is different in terms of operator ordering, depending on which method we use to perform the calculation.

```
Commutator[f[k, σ1] ** n_f[k, -σ1], H] /. σ1 → 1
```

$U ** f\dagger[k, -1] ** f[k, -1] ** f[k, 1] -$
$\quad U ** f\dagger[k, -1] ** n_f[k, 1] ** f[k, -1] ** f[k, 1] - \sum_j t[k, j] ** n_f[k, -1] ** f[j, 1] +$

$\quad \sum_i t[i, k] ** f\dagger[i, -1] ** f[k, -1] ** f[k, 1] - \sum_j t[k, j] ** f\dagger[k, -1] ** f[j, -1] ** f[k, 1]$

```
Commutator[f[k, 1] ** n_f[k, -1], H]
```

$U ** n_f[k, -1] ** f[k, 1] - \sum_j t[k, j] ** n_f[k, -1] ** f[j, 1] +$

$\quad \sum_i t[i, k] ** f\dagger[i, -1] ** f[k, -1] ** f[k, 1] - \sum_j t[k, j] ** f\dagger[k, -1] ** f[j, -1] ** f[k, 1]$

The two results are not identical. However, if we reorder the operators in the top output we replicate the bottom output, which has the operators correctly ordered.

```
StandardOrderQ[Commutator[f[k, σ1] ** n_f[k, -σ1], H] /. σ1 → 1]
```

$U ** n_f[k, -1] ** f[k, 1] - \sum_j t[k, j] ** n_f[k, -1] ** f[j, 1] +$

$\quad \sum_i t[i, k] ** f\dagger[i, -1] ** f[k, -1] ** f[k, 1] - \sum_j t[k, j] ** f\dagger[k, -1] ** f[j, -1] ** f[k, 1]$

Performing the same calculation with a spin down particle we obtain the same answer with the spins reversed.

```
Commutator[f[k, -1] ** n_f[k, 1], H]
```

$-U ** f†[k, 1] ** f[k, -1] ** f[k, 1] - \sum_i t[i, k] ** f†[i, 1] ** f[k, -1] ** f[k, 1] +$

$\sum_j t[k, j] ** f†[k, 1] ** f[j, -1] ** f[k, 1] + \sum_j t[k, j] ** f†[k, 1] ** f[k, -1] ** f[j, 1]$

## ◾ Problem 3 : $H f†_1^↑ f†_2^↓$ |Vacuum⟩

This demonstrates the properties of the Ket and Bra states, which need to be activated on the palette in addition to the Fermi, Bose or other relevant operators.

We can perform the product directly.

```
H ⊗ f†[1, 1] ** f†[2, -1] ** Ket[Vacuum]
```

$- \left(U \,\middle|\, \text{Vacuum}\right) ** f†[2, -1] ** f†[1, 1] ** n_f[2, 1] -$

$\left(U \,\middle|\, \text{Vacuum}\right) ** f†[1, -1] ** f†[2, -1] ** f†[1, 1] ** f[1, -1] +$

$\sum_i \left(\middle|\text{Vacuum}\right) t[i, 1]) ** f†[2, -1] ** f†[i, 1] + \sum_i \left(\middle|\text{Vacuum}\right) t[i, 2]) ** f†[i, -1] ** f†[1, 1] -$

$\sum_i \left(U \,\middle|\, \text{Vacuum}\right) ** f†[2, -1] ** f†[i, -1] ** f†[1, 1] ** f[i, 1] ** f[i, -1] ** f[i, 1] +$

$\sum_i \sum_j \sum_\sigma \left(\middle|\text{Vacuum}\right) t[i, j]) ** f†[2, -1] ** f†[1, 1] ** f†[i, \sigma] ** f[j, \sigma]$

We see that the destruction operators annihilate the vacuum state

```
f[i, σ] ** Ket[Vacuum]
```

0

We could use the ProductQ operator here, or equivalently use the symbol ⊗ between the operator and the ket.

```
ProductQ[H, Ket[Vacuum]]
```

0

```
Ket[1] = f†[1, 1] ** f†[2, -1] ** Ket[Vacuum]
```

f†[1, 1] ** f†[2, -1] ** |Vacuum⟩

```
ProductQ[H, f†[1, 1] ** f†[2, -1] ** Ket[Vacuum]]
```

$\sum_i t[i, 1] ** f†[2, -1] ** f†[i, 1] ** |\text{Vacuum}⟩ + \sum_i t[i, 2] ** f†[i, -1] ** f†[1, 1] ** |\text{Vacuum}⟩$

```
ProductQ[H, Ket[1]]
```

$\sum_i t[i, 1] ** f†[2, -1] ** f†[i, 1] ** |\text{Vacuum}⟩ + \sum_i t[i, 2] ** f†[i, -1] ** f†[1, 1] ** |\text{Vacuum}⟩$

```
H ⊗ f†[1, 1] ** f†[2, -1] ** Ket[Vacuum]
```

$\sum_i t[i, 1] ** f†[2, -1] ** f†[i, 1] ** |\text{Vacuum}⟩ + \sum_i t[i, 2] ** f†[i, -1] ** f†[1, 1] ** |\text{Vacuum}⟩$

We thereby see that all destruction operators have been killed and the correct answer is obtained.

Trying the above problem w/o ⊗,

```
H ** f†[1, 1] ** f†[2, -1] ** Ket[Vacuum]
```

$$\left( U \sum_i f†[i, 1] ** f[i, 1] ** f†[i, -1] ** f[i, -1] \right) ** f†[1, 1] ** f†[2, -1] ** \big| Vacuum \big\rangle +$$

$$\left( -\sum_i \sum_j \sum_\sigma f†[i, \sigma] ** f†[j, \sigma] ** t[i, j] \right) ** f†[1, 1] ** f†[2, -1] ** \big| Vacuum \big\rangle$$

Note that we would not get the correct answer if we just used NonCommutativeMultiply instead

of $\otimes$. Where it is appropriate to use NCM ($**$) rather than $\left(\bigotimes\right)$ might be confusing initially, but generally it is easy to deduce which one is appropriate. A rule of thumb is to use NCM in between single terms, and to use $\otimes$ in between larger expression involving many terms, Plus , Minus , or Sum functions).

■ **(IV A) Hubbard Model k - space Moment Calculations**

```
Clear[H]; H[l_, σ_, p_, q_, r_] :=
```
$$\sum_l \sum_\sigma \xi[l] f†[l, \sigma] ** f[l, \sigma] + (U/Ns) \sum_p \sum_q \sum_r f†[p+q, 1] ** f[p, 1] ** f†[r-q, -1] ** f[r, -1]$$

This is the Hubbard Hamiltonian in k space, with Ns as the number of sites, and the labels p,q,l, r as the wavevectors. and $\xi = \epsilon(k) - \mu$, with $\mu$ as the chemical potential and band dispersion $\epsilon(k)$. The dummy variables summed over in the right hand side (p,q,r,l,$\sigma$) are declared on the left hand side explicitly.

Throughout this notebook it is advisable to explicitly define the spin of the operators we are commuting with the Hamiltonian. This allows delta functions to be cancelled and thereby simplifies the result. Below we have computed the first moment using both spin up and spin down operators (1 for spin up, -1 for spin down). Also remember we must use new variables every time we commute with another Hamiltonian.

We now compute the symmetric moments of the Hubbard model- these are defined as $\omega_m = (-1)^m < \{[H, [H, ... [H, f], \text{f†}] > \text{with m nested commutators. Thus explicitly the first moment is the expectation: } \omega_1(k) = (-) < \{[H, f(k, \sigma)], \text{f†}(k, \sigma) > .$ Below we work out the operators that emerge from this nested commutator and the final anticommutator. (Clearly the expectation of these requires the solution of the eigensystem that we do not address).

A good reference for these moments is W. Nolting, Z. Physik 255, 25-- 39 (1972), his Eqs(8-11) correspond to our moments below after shifting $\omega_m = M^{(m+1)}$.

■ **First Moment**

```
-AntiCommutator[Commutator[H[l, σ, c, q, r], f[k, 1]], f†[k, 1]]
```

$$\sum_r \frac{U}{Ns} ** n_f[r, -1] + \xi[k]$$

■ **Second Moment**

```
AntiCommutator[
 Commutator[H[l₁, σ₁, χ, θ₁, ρ₁], Commutator[H[l, σ, c, q, r], f[k, 1]]], f†[k, 1]]
```

$$\sum_r \frac{U\,\xi[k]}{Ns} ** n_f[r, -1] + \sum_{\rho_1} \frac{U\,\xi[k]}{Ns} ** n_f[\rho_1, -1] + \sum_{\theta_1} \sum_{\rho_1} \frac{U^2}{Ns^2} ** n_f[\rho_1, -1] -$$

$$\sum_r \sum_{\theta_1} \sum_{\rho_1} \frac{U^2}{Ns^2} ** f†[r+\theta_1, -1] ** f†[-\theta_1+\rho_1, -1] ** f[r, -1] ** f[\rho_1, -1] -$$

$$\sum_r \sum_{\theta_1} \sum_{\rho_1} \frac{U^2}{Ns^2} ** f†[-\theta_1+\rho_1, -1] ** f†[k+\theta_1, 1] ** f[r, -1] ** f[k-r+\rho_1, 1] +$$

$$\sum_q \sum_{\theta_1} \sum_{\rho_1} \frac{U^2}{Ns^2} ** f†[-q-\theta_1+\rho_1, -1] ** f†[k+\theta_1, 1] ** f[\rho_1, -1] ** f[k-q, 1] + \xi[k]^2$$

Clearly a relabeling of variables would help combine terms here; it is left as an exercise for the reader.

■ **Third Moment**

The third moment can be calculated, but it requires further simplification. Due to the length of the result it is not shown below. Remove the semicolon at the end to see the large output. This calculation is long, and it takes around ten seconds to complete.

```
-AntiCommutator[Commutator[H[λ, ß, ξ, ω, ι],
    Commutator[H[l₁, σ₁, χ, θ₁, ρ₁], Commutator[H[l, σ, p, q, r], f[k, 1]]]], f†[k, 1]];
```

■ **(IV B) Hubbard Model r - space Moment Calculations**

Here the same calculation as above can be done in real space-finally taking Fourier transforms over the two fixed site indices r1,r2 with wave vector k gives the moments $\omega_m(k)$. The results are equivalent to those in $k$ space shown

above. It is interesting to see these since the Pauli principle identities such as $n ** n = n$ are utilized here and we can finally take Fourier transforms and sum over the fixed external sites (r1 , r2) .

```
Clear[H];
H[j_] := (-1) * ∑   ∑   ∑   t_{k[j],m[j]} f†[k[j], u[j]] ** f[m[j], u[j]] -
           k[j] m[j] u[j]

    μ ∑   ∑   f†[s[j], v[j]] ** f[s[j], v[j]] - U ∑   n_f[t[j], 1] ** n_f[t[j], -1]
       s[j] v[j]                                    t[j]
```

■ **First Moment**

```
-AntiCommutator[Commutator[H[1], f[r1, 1]], f†[r2, 1]]
```

$- (U\, \delta[\text{r1}, \text{r2}]) ** n_f[\text{r1}, -1] - t_{\text{r1,r2}} - \mu\, \delta[\text{r1}, \text{r2}]$

■ **Second Moment**

```
AntiCommutator[Fold[Commutator, f[r1, 1], {H[1], H[2]}], f†[r2, 1]]
```

$(U\, t_{\text{r1,r2}}) ** n_f[\text{r1}, -1] + (U\, t_{\text{r1,r2}}) ** n_f[\text{r2}, -1] + \left(U^2\, \delta[\text{r1}, \text{r2}]\right) ** n_f[\text{r1}, -1] +$

$2\ (U\, \mu\, \delta[\text{r1}, \text{r2}]) ** n_f[\text{r1}, -1] + 2\, \mu\, t_{\text{r1,r2}} + \sum_{m[2]} (U\, t_{\text{r1,m[2]}}\, \delta[\text{r1}, \text{r2}]) ** f†[\text{r1}, -1] ** f[m[2], -1] -$

$\sum_{k[2]} (U\, t_{k[2],\text{r1}}\, \delta[\text{r1}, \text{r2}]) ** f†[k[2], -1] ** f[\text{r1}, -1] + \sum_{m[1]} t_{\text{r1,m[1]}}\, t_{m[1],\text{r2}} + \mu^2\, \delta[\text{r1}, \text{r2}]$

■ **Third Moment**

(* Warning: Produces a large output *)

```
-AntiCommutator[Fold[Commutator, f[r1, 1], {H[1], H[2], H[3]}], f†[r2, 1]];
```

■ **(V) Spins at work: Eight Vertex Model Yang - Baxter Equation**

● **Activate the Pauli matrix symbol "σ"**

We use DiracQ to simplify some algebra that arises in the well known work of Baxter on the two dimensional 8 vertex model (R. J. Baxter, Ann. Phys. vol 76, p1 (1973)) . The transfer matrix of the model with a parameter u is written as:

$$T(u) = \text{Tr}_g\left[L_{N,g}(u)\, L_{N-1,g}(u) \dots L_{1,g}(u)\right], \tag{1}$$

where each of the objects $L_{n,g}(u)$ is $a$ scattering matrix

$$L_{n,g}(u) = \frac{a+b}{2} + \frac{a-b}{2}\, \sigma_n^z \sigma_g^z + \frac{c+d}{2}\, \sigma_n^x \sigma_g^x + \frac{c-d}{2}\, \sigma_n^y \sigma_g^y. \tag{2}$$

Here the Pauli matrices are defined for sites n=1,N and an extra "ghost" site "g". The Boltzmann weights a, b, c, and d are functions of the spectral parameter u. The fundamental commutation relation of the transfer matrices [T(u),T(v)]=0 defines the integrability of this model and also is the key to its solution. Baxter shows that this commuting transfer matrix relation is satisfied if the local relation

$L_{n,g_2}[u] \otimes L_{n,g_1}[u'] \otimes R_{g_1,g_2}[u''] = R_{g_1,g_2}[u''] \otimes L_{n,g_1}[u'] \otimes L_{n,g_2}[u]$ (Baxter's equation)

is satisfied, where explicitly

$$L_{n,g_2}(u) = \frac{a+b}{2} + \frac{a-b}{2} \sigma_{g_2}^z \sigma_n^z + \frac{c+d}{2} \sigma_{g_2}^x \sigma_n^x + \frac{c-d}{2} \sigma_{g_2}^y \sigma_n^y$$

$$L_{n,g_1}(u') = \frac{a'+b'}{2} + \frac{a'-b'}{2} \sigma_{g_1}^z \sigma_n^z + \frac{c'+d'}{2} \sigma_{g_1}^x \sigma_n^x + \frac{c'-d'}{2} \sigma_{g_1}^y \sigma_n^y$$

$$R_{g_2,g_1}(u'') = \frac{a''+b''}{2} + \frac{a''-b''}{2} \sigma_{g_1}^z \sigma_{g_2}^z + \frac{c''+d''}{2} \sigma_{g_1}^x \sigma_{g_2}^x + \frac{c''-d''}{2} \sigma_{g_1}^y \sigma_{g_2}^y \, .$$

(3)

Let us use the functions of the DiracQ package to see if we can find the requirements on the u's such that (10) is satisfied. For convenience let us set the three distinct sites in the problem explicitly as: n=3, $g_2 = 1$, $g_1 = 2$. This allows the code to automatically solve the Kronecker $\delta$ functions, as shown below. For more information on use of the Kronecker $\delta$ see the Tutorial.

**δ[g₁, g₂]**

$\delta[g_1, g_2]$

**δ[2, 1]**

0

**n = 3; g₂ = 1; g₁ = 2;**

**L_{n,g₂}[W] = $\frac{a+b}{2}$ + $\frac{a-b}{2}$ σ[g₂, z] ** σ[n, z] + $\frac{c+d}{2}$ σ[g₂, x] ** σ[n, x] + $\frac{c-d}{2}$ σ[g₂, y] ** σ[n, y];**

**L_{n,g₁}[W'] =**
**$\frac{a'+b'}{2}$ + $\frac{a'-b'}{2}$ σ[g₁, z] ** σ[n, z] + $\frac{c'+d'}{2}$ σ[g₁, x] ** σ[n, x] + $\frac{c'-d'}{2}$ σ[g₁, y] ** σ[n, y];**

**R_{g₁,g₂}[W''] = $\frac{a''+b''}{2}$ + $\frac{a''-b''}{2}$ σ[g₁, z] ** σ[g₂, z] +**
**$\frac{c''+d''}{2}$ σ[g₁, x] ** σ[g₂, x] + $\frac{c''-d''}{2}$ σ[g₁, y] ** σ[g₂, y];**

Let us first compute the product of the LHS Baxter's equation. For this we will use the function, which will apply the algebra of Pauli matrices and collect c terms. Because there are 8 terms in each scattering matrix the number of terms on both the right and left hand sides of equation is large (128 terms in all). Because of this we will suppress the output.

**LHS = L_{n,g₂}[W] ⊗ (L_{n,g₁}[W'] ⊗ R_{g₁,g₂}[W'']);**

**RHS = R_{g₁,g₂}[W''] ⊗ (L_{n,g₁}[W'] ⊗ L_{n,g₂}[W]);**

**{Length[LHS], Length[RHS]}**

{128, 128}

We can take a peek at the output:

**Short[LHS, 3]**

$\frac{1}{4}$ (a a′ a″) ** σ[1, z] ** σ[2, z] + $\frac{1}{4}$ (a a′ a″) ** σ[1, z] ** σ[3, z] +

$\frac{1}{4}$ (a a′ a″) ** σ[2, z] ** σ[3, z] + ≪170≫ + $\frac{1}{4}$ d d′ c″ + $\frac{1}{4}$ d c′ d″ + $\frac{1}{4}$ c d′ d″

**Short[RHS, 3]**

$\frac{1}{4}$ (a a′ a″) ** σ[1, z] ** σ[2, z] + $\frac{1}{4}$ (a a′ a″) ** σ[1, z] ** σ[3, z] +

$\frac{1}{4}$ (a a′ a″) ** σ[2, z] ** σ[3, z] + ≪170≫ + $\frac{1}{4}$ d d′ c″ + $\frac{1}{4}$ d c′ d″ + $\frac{1}{4}$ c d′ d″

We now need to collect terms which involve the same set of operators. We can then set the sum of the coefficients of all the terms with identical operators to zero. We will now use the function TakeQPart, to extract the string of operators from each term. By using the Union function we can also remove duplicate expressions. We are thereby left with a list of all the different strings of operators that appear anywhere throughout the expression.

**OperatorTerms = Union[TakeQPart[LHS - RHS]]**

{σ[1, x] ** σ[2, y] ** σ[3, z], σ[1, x] ** σ[2, z] ** σ[3, y], σ[1, y] ** σ[2, x] ** σ[3, z],
 σ[1, y] ** σ[2, z] ** σ[3, x], σ[1, z] ** σ[2, x] ** σ[3, y], σ[1, z] ** σ[2, y] ** σ[3, x]}

**Length[OperatorTerms]**

6

We can now use the QCoefficient function to find the coefficients of a string of operators. The use of this function is demonstrated below.

**QCoefficient[LHS - RHS, OperatorTerms[[1]]]**

$\frac{1}{2}$ i d c′ a″ − $\frac{1}{2}$ i c d′ a″ + $\frac{1}{2}$ i c c′ b″ − $\frac{1}{2}$ i d d′ b″ + $\frac{1}{2}$ i b a′ c″ − $\frac{1}{2}$ i a b′ c″ + $\frac{1}{2}$ i a a′ d″ − $\frac{1}{2}$ i b b′ d″

The function has found all of the terms with operators that match the pattern of OperatorTerms[[1]] and summed their coefficients. This term must go to zero for the Yang - Baxter relation to hold. We can work this function into a loop and find all such expressions at once, as shown below.

**Do[expression[n] = QCoefficient[LHS - RHS, OperatorTerms[[n]]];**
 **Print[expression[n] == 0], {n, Length[OperatorTerms]}]**

$\frac{1}{2}$ i d c′ a″ − $\frac{1}{2}$ i c d′ a″ + $\frac{1}{2}$ i c c′ b″ − $\frac{1}{2}$ i d d′ b″ + $\frac{1}{2}$ i b a′ c″ − $\frac{1}{2}$ i a b′ c″ + $\frac{1}{2}$ i a a′ d″ − $\frac{1}{2}$ i b b′ d″ == 0

− $\frac{1}{2}$ i d a′ a″ + $\frac{1}{2}$ i c b′ a″ − $\frac{1}{2}$ i c a′ b″ + $\frac{1}{2}$ i d b′ b″ − $\frac{1}{2}$ i b c′ c″ − $\frac{1}{2}$ i a d′ c″ − $\frac{1}{2}$ i a c′ d″ + $\frac{1}{2}$ i b d′ d″ == 0

$\frac{1}{2}$ i d c′ a″ − $\frac{1}{2}$ i c d′ a″ − $\frac{1}{2}$ i c c′ b″ + $\frac{1}{2}$ i d d′ b″ − $\frac{1}{2}$ i b a′ c″ + $\frac{1}{2}$ i a b′ c″ + $\frac{1}{2}$ i a a′ d″ − $\frac{1}{2}$ i b b′ d″ == 0

− $\frac{1}{2}$ i d a′ a″ − $\frac{1}{2}$ i c b′ a″ + $\frac{1}{2}$ i c a′ b″ + $\frac{1}{2}$ i d b′ b″ + $\frac{1}{2}$ i b c′ c″ + $\frac{1}{2}$ i a d′ c″ − $\frac{1}{2}$ i a c′ d″ − $\frac{1}{2}$ i b d′ d″ == 0

$\frac{1}{2}$ i a c′ a″ − $\frac{1}{2}$ i b d′ a″ − $\frac{1}{2}$ i b c′ b″ + $\frac{1}{2}$ i a d′ b″ − $\frac{1}{2}$ i c a′ c″ + $\frac{1}{2}$ i d b′ c″ + $\frac{1}{2}$ i d a′ d″ − $\frac{1}{2}$ i c b′ d″ == 0

− $\frac{1}{2}$ i a c′ a″ − $\frac{1}{2}$ i b d′ a″ + $\frac{1}{2}$ i b c′ b″ + $\frac{1}{2}$ i a d′ b″ + $\frac{1}{2}$ i c a′ c″ + $\frac{1}{2}$ i d b′ c″ − $\frac{1}{2}$ i d a′ d″ − $\frac{1}{2}$ i c b′ d″ == 0

These are the final equations. By recombining them and simplifying we reproduce Baxter's equations.

```
Expand[(expression[2] + expression[4]) / ⅈ] == 0
Expand[(expression[1] + expression[3]) / ⅈ] == 0
Expand[(expression[5] + expression[6]) / ⅈ] == 0
Expand[(expression[2] - expression[4]) / ⅈ] == 0
Expand[(expression[1] - expression[3]) / ⅈ] == 0
Expand[(expression[5] - expression[6]) / ⅈ] == 0
```

$-d\,a'\,a'' + d\,b'\,b'' + a\,d'\,c'' - a\,c'\,d'' == 0$

$d\,c'\,a'' - c\,d'\,a'' + a\,a'\,d'' - b\,b'\,d'' == 0$

$-b\,d'\,a'' + a\,d'\,b'' + d\,b'\,c'' - c\,b'\,d'' == 0$

$c\,b'\,a'' - c\,a'\,b'' - b\,c'\,c'' + b\,d'\,d'' == 0$

$c\,c'\,b'' - d\,d'\,b'' + b\,a'\,c'' - a\,b'\,c'' == 0$

$a\,c'\,a'' - b\,c'\,b'' - c\,a'\,c'' + d\,a'\,d'' == 0$

These are six homogeneous equations relating the twelve parameters (a,b,c,d) etc. Baxter proceeds to give an elegant analysis of these by parameterization in terms of the spectral parameters u, u', u''. The solution is well described in the original paper; we see above that DiracQ helps in automating the tedious calculations.

## ■ (VI) 1D Hubbard Model SSS Matrix Problem

### ● Activate the Pauli matrix symbol "σ"

We next demonstrate the use of DiracQ in the context of the 1-d Hubbard model. It satisfies the Yang Baxter relation, analogous to the 8 vertex model discussed in the previous example, and the proof of this is given in B. S. Shastry, Phys. Rev. Letts. vol 56, 1529, 2453 (1986). A relatively simple analytical proof is also to be found in B. S. Shastry, J. Stat. Phys. vol 50, 57 (1988), using the decorated star triangle relations. We recall this result before proceeding further. It may be noted that the R matrix of this model, defined below, is currently of great interest in connection with the problem of the ADS/CFT correspondence for N = 4 SUSY Yang-Mills theory in four dimensions following the work of N. Beisert, J.Stat.Mech. 0701 (2007) P01017, arXiv:nlin/0610017.

The 1-dimensional Hubbard model Hamiltonian using the Jordan Wigner transformation and with suitable boundary conditions can be written in terms of Pauli spin matrices as

$$H = \sum_n H_{n+1,n}; \quad H_{n+1,n} = (\sigma_n^+ \, \sigma_{n+1}^- + \sigma_{n+1}^+ \, \sigma_n^-) + (\tau_n^+ \, \tau_{n+1}^- + \tau_{n+1}^+ \, \tau_n^-) + \frac{1}{4} \, U \, \sigma_n^z \, \tau_n^z \, . \tag{4}$$

This corresponds to the quantum Hamiltonian, commuting with the transfer matrix of two Free fermi models coupled in a specific way. We start with a single Pauli species scattering matrix :

$$l_{ij}^{(\sigma)}(\theta) = \frac{a+b}{2} + \frac{a-b}{2} \, \sigma_i^z \sigma_j^z + \frac{c}{2} \left( \sigma_i^x \, \sigma_j^x + \sigma_i^y \, \sigma_j^y \right) \tag{5}$$

Here the parameters a b c satisfy the Free Fermi condition $a^2 + b^2 = c^2$. This is conveniently parametrized by setting $a = c \, \text{Cos}(\theta)$ $b = c \, \text{Sin}(\theta)$. For the two sites 1 and 2 we define $a_1 = \text{Cos}[\theta_1]$, $b_1 = \text{Sin}[\theta_1]$, $a_2 = \text{Cos}[\theta_2]$, $b_2 = \text{Sin}[\theta_2]$ and denote $\left( W_j \equiv (\theta_j, \, h_j) \right)$.

We may write two sets of scattering matrices $l_{ij}^{(\sigma)}$ and $l_{ij}^{(\tau)}$ for the two species at the same $\theta$ :

$$l_{ij}(\theta) = l_{ij}^{(\sigma)}(\theta) \otimes l_{ij}^{(\tau)}(\theta) \tag{6}$$

The scattering operator for the Hubbard model is constucted as

$$L_{3n}(W_n) = l_{3n}(\theta_n) \, I_n(h_n),$$

The transfer matrix can now be written as

$$T(W) = \underset{o}{\text{Tr}}[L_{No}(W) \, L_{N-1 \, o}(W) \, ... \, L_{1 \, o}(W)]. \tag{7}$$

With these definitions, the transfer matrix forms a commuting family $[T(W_1), \, T(W_2)] = 0$ provided the scattering matrix is proven to satisfy the Yang − Baxter equation $(W_n \equiv (\theta_n, h_n))$

$$L_{31}(W_1) \, L_{32}(W_2) \, S_{12}(W_2 \mid W_1) = S_{12}(W_2 \mid W_1) \, L_{32}(W_2) \, L_{31}(W_1). \tag{8}$$

The S operator found by Shastry is given by:

$$S_{12}(W_2 \mid W_1) = \rho[\cos(\theta_2 + \theta_1) \cos(h_2 - h_1) \, l_{12}(\theta_2 - \theta_1) + \cos(\theta_2 - \theta_1) \sinh(h_2 - h_1) \, l_{12}(\theta_2 + \theta_1) \, \sigma_2^z \, \tau_2^z]. \tag{9}$$

An extra and crucial sufficiency condition is that the fields $h_n$ (n = 1, 2) must satisfy

$$\frac{\text{Sinh}[2 \, h_n]}{a_n \, b_n} = \frac{U}{2} \, , \tag{10}$$

where U is the interaction term in the Hubbard Hamiltonian. The peculiarity of this model is that $S(W_2 \mid W_1)$ depends on the spectral parameters $W_1$ and $W_2$ separately, and cannot be written in terms of the difference of the spectral parameters. This relation was proved by Shastry analytically using the decorated star triangle relations.

A more (technically) ambitious question is to ask is if we can construct a more general inhomogeneous model than the Hubbard model, with a transfer matrix T (W | $W_1$, ... $W_n$) depending on the N fixed parameters $W_j$

$$T = \text{Tr}_o[S_{No}(W_1 \mid W_2) \, S_{N-1 \, o}(W \mid W_{N-1}) \ldots S_{1 \, o}(W \mid W_1)]. \tag{11}$$

so that on setting $W_j$ to 0, we get back the Hubbard transfer matrix above. The inhomogeneous matrix T commutes for different weights according to

$$[T(W \mid W_1 \ldots W_N), \, T'(W' \mid (W_1 \ldots W_N)] = 0. \tag{12}$$

The sufficiency condition for this is the more general Yang Baxter relation

$$S_{31}(W_1 \mid W_3) \, S_{32}(W_2 \mid W_3) \, S_{12}(W_2 \mid W_1) = S_{12}(W_2 \mid W_1) \, S_{32}(W_2 \mid W_3) \, S_{31}(W_1 \mid W_3). \tag{13}$$

This is Eq( 4.14) of the paper and several typical matrix elements were verified by Shastry by explicit calculation, who conjectured that this equation is true for all matrix elements. Since the number of terms is very large (see below), an exhaustive enumerative proof was not given. We will provide such an enumerative proof using DiracQ next.

For non zero $W_3$, an analytical proof was actually found much later in the work of M. Shiroishi and M. Wadati - Journal of the Physical Society of Japan, Vol 64, 57-63 (1995), and thus the following enumerative proof is not the first proof. However, it does demonstrate the power of DiracQ quite well. The large number of terms involved make the algebra prohibitive by a hand calculation for most of us! Using the package however we are able to compute the product of three S operators in a number of minutes.

The first step is to input the form for the scattering matrix and for the S operators. Here, we denote the two separate species of Pauli matrices not by use of another symbol, but rather by using a third index to denote the species of the operator, as shown below.

$$\sigma_1^z \to \sigma[1, z, 1] \,, \; \tau_1^z \to \sigma[1, z, 2] \tag{14}$$

In this notation the scattering matrix and the S operator are written below as general functions.

```
Clear[A, l, S];
l_{u_,v_}[n_, p_, q_] :=
    (A[n, 1, p, q] + A[n, -1, p, q] σ[v, z, 1] ** σ[u, z, 1] + σ[v, x, 1] ** σ[u, x, 1] +
        σ[v, y, 1] ** σ[u, y, 1]) ⊗ (A[n, 1, p, q] + A[n, -1, p, q] σ[v, z, 2] ** σ[u, z, 2] +
        σ[v, x, 2] ** σ[u, x, 2] + σ[v, y, 2] ** σ[u, y, 2]);
S_{u_,v_} := Cos[θ_v + θ_u] Cosh[h_v - h_u] l_{u,v}[-1, θ_v, θ_u] +
    Cos[θ_v - θ_u] Sinh[h_v - h_u] l_{u,v}[1, θ_v, θ_u] σ[v, z, 1] σ[v, z, 2]
```

The Boltzmann weights in the scattering matrix are not written explicitly because this increases the number of terms involved in the product and their definition will not be relevant until after the S operator products have been calculated. They are therefore represented by the function A, which will be defined after the product of the three S operators on each side of (36) has been computed. We will set

```
rulesimplify = { h_i_ -> 1 / 2 ArcSinh[U / 2 Cos[θ_i] Sin[θ_i]] ,
    A[n_, m_, p_, q_] -> Cos[p + n q] + m Sin[p + n q] }
```

after the calculation with the operators is performed, and it is easy to see that this replacement reproduces the scattering operator above with a scale factor of 2 which cancels out. Here we compute the product of three S , matrices. Each S matrix contains about thirty terms, we we wexpect the product to contain on the order of 27,000 terms. This expression is so large as to be prohibitive to compute by hand.

```
{time, LHS} = Timing[S_{3,1} ⊗ (S_{3,2} ⊗ S_{1,2})];

time / 60

2.79112
```

The timing function informs us that this operation takes around three minutes on our testing machine.

**Length[LHS]**

27 360

We see that there are approximately 32,000 terms in the final product. Below an arbitrary example of a single term is shown. We expect similar timing and length for the RHS.

**LHS[[6]]**

$A[-1, 1, \Theta_1, \Theta_3]^2 A[-1, 1, \Theta_2, \Theta_1]^2 A[-1, 1, \Theta_2, \Theta_3]^2$
$Cos[\Theta_1 + \Theta_2] Cos[\Theta_1 + \Theta_3] Cos[\Theta_2 + \Theta_3] Cosh[h_1 - h_2] Cosh[h_1 - h_3] Cosh[h_2 - h_3]$

We now compute the RHS

**RHS = $S_{1,2} \otimes (S_{3,2} \otimes S_{3,1})$;**

**Length[RHS]**

27 360

We have now computed both the left and right hand sides of equation 13. If the equation holds, they are equal. We verify this by subtracting the two and returnign zero.

**subtraction = LHS - RHS;**

**Length[subtraction]**

13 472

We would have hoped that when we subtracted the LHS from the RHS that the result was zero. It appears we will have to do more manipulation to show that the LHS is equal to the RHS. We now have a very large expression, and we need to manipulate it in several ways. We need to find all distinct strings of operators and then find the coefficient of all of each string. We must then show that every one of these coefficients vanishes. DiracQ contains functions that will perform the first two. However, were we to use the functions outright the computation time would be very long. Every expression input into a DiracQ function is first organized into lists that can be manipulated (see the ' Brief Explanation of Form' section). For a large expression such as the one we are working with, organizing the expression takes a very long time. The expression must then be recompiled into and understandable form. We can circumvent some of these steps by organizing an expression once and specifiying to subsequent functions that the input has already been organized and that this step should be skipped. This is done by specifying' Organized Expression -> True' as an option. This process is demonstrated below.

**? Organize**

Organize is the function that enables the DiracQ package to understand user input. Organize takes a mathematical expression as input and yields a nested list that contains the atoms of the input ordered according to their properties. Numbers, summed indices, c numbers, and q numbers are separated into groups. Each term of the input separated by plus sign constitutes a separate list of items in the output. Example:

Organize[(⊞) $\sum_{index}$ (c ⊞)∗(q ⊞)]

={{⊞,{index},{c ⊞},{q ⊞}}}

Organize[(⊞$_1$) $\sum_{index_1}$ (c ⊞$_1$)∗(q ⊞$_1$)+(⊞$_2$) $\sum_{index_2}$ (c ⊞$_2$)∗(q ⊞$_2$)]

={{⊞$_1$,{index$_1$},{c ⊞$_1$},{q ⊞$_1$}},{⊞$_2$,{index$_2$},{c ⊞$_2$},{q ⊞$_2$}}}
For a more in depth explanation see the DiracQ writeup notebook.

**{time, organizedsubtraction} = Timing[Organize[subtraction]];**

**time / 60**

1.37672

We see that on our test machine the Organize function takes a few minutes to evaluate and is computationally quite expensive. Below we use the TakeQPart function to find the strings of operators that appear in OrganizedSubtraction, specifying that the input is already an organized expression to save some computational time.

```
OperatorTerms = Union[TakeQPart[organizedsubtraction, OrganizedExpression → True]];
```

```
Length[OperatorTerms]
```

```
472
```

```
OperatorTerms[[300]]
```

$\sigma[1, y, 1] ** \sigma[2, z, 1] ** \sigma[3, x, 1] ** \sigma[2, z, 2] ** \sigma[3, z, 2]$

The above output shows that there are 472 distinct strings of operators found in the expression and an example of one such string. We can now use the QCoefficient function to find the coefficients of a string of operators.

```
? QCoefficient
```

QCoefficient[expression,form] will scan the expression for terms containing a string of operators that match 'form'. The function will output the coefficient of the operator(s) specified by 'form'. Only exact matches of the string of operators are found. If several terms are found in expression containing terms that match 'form' the output will be a sum of the coefficients of the specified operators.

The result is not printed due to the length and complexity, though we do show the first term in the expression. See the appendix section, or remove the semicolon at the end to see the expression.

```
sum[1] =
    QCoefficient[organizedsubtraction, OperatorTerms[[1]], OrganizedExpression → True];
```

```
Length[sum[1]]
```

```
24
```

```
sum[1][[1]]
```

$4 A[-1, -1, \theta_2, \theta_3] A[1, -1, \theta_1, \theta_3] \mathrm{Cos}[\theta_1 - \theta_2]$
$\mathrm{Cos}[\theta_1 - \theta_3] \mathrm{Cos}[\theta_2 + \theta_3] \mathrm{Cosh}[h_2 - h_3] \mathrm{Sinh}[h_1 - h_2] \mathrm{Sinh}[h_1 - h_3]$

We now apply the definition of the Boltzmann weights.

```
rulesimplify =
    {h_i_ -> 1 / 2 ArcSinh[U / 2 Cos[θ_i] Sin[θ_i]], A[n_, m_, p_, q_] -> Cos[p + n q] + m Sin[p + n q]};
```

```
Simplify[sum[1] /. rulesimplify]
```

```
0
```

We see that the coefficients of this one string of operators simplify to zero. We must show that each such term vanishes for all 472 strings of operators. The command below will evaluate each such term.

```
Do[sum[i] = Simplify[QCoefficient[organizedsubtraction,
    OperatorTerms[[i]], OrganizedExpression → True] /. rulesimplify], {i, 1, 472}]
```

```
Table[sum[i], {i, 1, 472}]
```

{0, 0, 0, 0, 0, 0, 0, 0, 0, 0, 0, 0, 0, 0, 0, 0, 0, 0, 0, 0, 0, 0, 0, 0, 0, 0, 0, 0, 0, 0, 0, 0, 0, 0,
0, 0, 0, 0, 0, 0, 0, 0, 0, 0, 0, 0, 0, 0, 0, 0, 0, 0, 0, 0, 0, 0, 0, 0, 0, 0, 0, 0, 0, 0, 0, 0, 0, 0,
0, 0, 0, 0, 0, 0, 0, 0, 0, 0, 0, 0, 0, 0, 0, 0, 0, 0, 0, 0, 0, 0, 0, 0, 0, 0, 0, 0, 0, 0, 0, 0, 0, 0,
0, 0, 0, 0, 0, 0, 0, 0, 0, 0, 0, 0, 0, 0, 0, 0, 0, 0, 0, 0, 0, 0, 0, 0, 0, 0, 0, 0, 0, 0, 0, 0, 0, 0,
0, 0, 0, 0, 0, 0, 0, 0, 0, 0, 0, 0, 0, 0, 0, 0, 0, 0, 0, 0, 0, 0, 0, 0, 0, 0, 0, 0, 0, 0, 0, 0, 0, 0,
0, 0, 0, 0, 0, 0, 0, 0, 0, 0, 0, 0, 0, 0, 0, 0, 0, 0, 0, 0, 0, 0, 0, 0, 0, 0, 0, 0, 0, 0, 0, 0, 0, 0,
0, 0, 0, 0, 0, 0, 0, 0, 0, 0, 0, 0, 0, 0, 0, 0, 0, 0, 0, 0, 0, 0, 0, 0, 0, 0, 0, 0, 0, 0, 0, 0, 0, 0,
0, 0, 0, 0, 0, 0, 0, 0, 0, 0, 0, 0, 0, 0, 0, 0, 0, 0, 0, 0, 0, 0, 0, 0, 0, 0, 0, 0, 0, 0, 0, 0, 0, 0,
0, 0, 0, 0, 0, 0, 0, 0, 0, 0, 0, 0, 0, 0, 0, 0, 0, 0, 0, 0, 0, 0, 0, 0, 0, 0, 0, 0, 0, 0, 0, 0, 0, 0,
0, 0, 0, 0, 0, 0, 0, 0, 0, 0, 0, 0, 0, 0, 0, 0, 0, 0, 0, 0, 0, 0, 0, 0, 0, 0, 0, 0, 0, 0, 0, 0, 0, 0,
0, 0, 0, 0, 0, 0, 0, 0, 0, 0, 0, 0, 0, 0, 0, 0, 0, 0, 0, 0, 0, 0, 0, 0, 0, 0, 0, 0, 0, 0, 0, 0, 0, 0,
0, 0, 0, 0, 0, 0, 0, 0, 0, 0, 0, 0, 0, 0, 0, 0, 0, 0, 0, 0, 0, 0, 0, 0, 0, 0, 0, 0, 0, 0, 0, 0, 0, 0,
0, 0, 0, 0, 0, 0, 0, 0, 0, 0, 0, 0, 0, 0, 0, 0, 0, 0, 0, 0, 0, 0, 0, 0, 0, 0, 0, 0, 0, 0, 0, 0, 0, 0,
0, 0, 0, 0, 0, 0, 0, 0, 0, 0, 0, 0, 0, 0, 0, 0, 0, 0, 0, 0, 0, 0, 0, 0, 0, 0, 0, 0, 0, 0, 0, 0, 0, 0}

Thus we see that every term simplifies to zero and thus proves the result.

■ **(VII) Hubbard Model Currents**

Below we define the Hamiltonian of the Hubbard model using fermionic operators. The system is taken to have periodic boundary conditions such that (N + a) a where N is the number of sites. We also write two currents which have been shown to commute with the Hamiltonian in the papers B.S.Shastry Phys. Rev. Letts. vol 56, 1529, 2453 (1986) and especially J.Stat.Phys, vol 50, 57 (1988). Here it is shown that for a system with odd number of sites we have a conservation law termed $j_A$ (defined below) that stands apart from the other conservation laws. A current $j_B$ is also found (defined below) and unlike $j_A$ generalizes to many more currents since it is the first non trivial member of the currents contained in the transfer matrix. Note that $j_A$ corresponds to infinite ranged hopping of the doubly occupied sites adding to short ranged hops of fermions, whereas $j_B$ corresponds to second neighbor hopping.

We will now verify the commutation of $j_A$ for odd number of sites using DiracQ and the commutation of $j_B$ for both odd and even sites. Notice that we write the Hamiltonian as well as the currents as functions of the number of sites NN (N is a predefined symbol in *Mathematica*). This will allow us to vary the number of sites between even and odd values.

```
Clear[H]; Remove[j];
```

$$H[NN\_] := -t \sum_{n=1}^{NN} Sum[(f\dagger[n+1, \sigma] ** f[n, \sigma] + f\dagger[n, \sigma] ** f[n+1, \sigma]), \{\sigma, -1, 1, 2\}] +$$

$$U \sum_{m=1}^{NN} n_f[m, 1] ** n_f[m, -1]$$

$$j_A[NN\_] := i\, t \sum_{l=1}^{NN} Sum[(-f\dagger[1, \sigma] ** f[1+1, \sigma] + f\dagger[1+1, \sigma] ** f[1, \sigma]), \{\sigma, -1, 1, 2\}] +$$

$$i\, U \sum_{r=1}^{NN} \sum_{l=1}^{NN-1} (-1)^l\, f\dagger[1+r, 1] ** f\dagger[1+r, -1] ** f[r, -1] ** f[r, 1];$$

$$j_B[NN\_] := i\, t \sum_{o=1}^{NN} Sum[f\dagger[o+2, \sigma] ** f[o, \sigma] - f\dagger[o, \sigma] ** f[o+2, \sigma], \{\sigma, -1, 1, 2\}] +$$

$$i\, U \sum_{o=1}^{NN} Sum[f\dagger[o+1, \sigma] ** f[o, \sigma] - f\dagger[o, \sigma] ** f[o+1, \sigma], \{\sigma, -1, 1, 2\}] +$$

$$i\, U \sum_{o=1}^{NN} Sum[(f\dagger[o, \sigma] ** (f[o+1, \sigma] - f[o-1, \sigma]) - (f\dagger[o+1, \sigma] - f\dagger[o-1, \sigma]) ** f[o, \sigma])$$
$$n_f[o, -\sigma], \{\sigma, -1, 1, 2\}];$$

First we will perform the commutator of the Hamiltonian with $j_A$ for a system with 5 sites. As the number of sites increases so does the number of operations performed,

and therefore we will only perform these commutators for relatively small systems. Even so the computing time is nonnegligible. We must include the periodic boundary conditions of our system. These are written as replacement rules. The replacement rules must be specified to each term, and cannot be specified at the end of computation, as this will lead to an erroneous answer.

```
PBCrule[n_] := {f[i_, a_] /; i > n → f[i - n, a], f[i_, a_] /; i < 1 → f[i + n, a],
    f†[i_, a_] /; i > n → f†[i - n, a], f†[i_, a_] /; i < 1 → f†[i + n, a]}
```

```
Timing[Commutator[H[5] /. PBCrule[5], jA[5] /. PBCrule[5]]]
```

{20.779, 0}

Performing the same computation with a four site sytem we see that the result is nonvanishing, as we expected.

```
Commutator[H[4] /. PBCrule[4], jA[4] /. PBCrule[4]]
```

```
2 i (t U) ** f†[1, -1] ** f†[1, 1] ** f[1, -1] ** f[2, 1] +
  2 i (t U) ** f†[1, -1] ** f†[1, 1] ** f[2, -1] ** f[1, 1] -
  2 i (t U) ** f†[1, -1] ** f†[2, 1] ** f[2, -1] ** f[2, 1] -
  2 i (t U) ** f†[1, -1] ** f†[4, 1] ** f[1, -1] ** f[1, 1] -
  2 i (t U) ** f†[2, -1] ** f†[1, 1] ** f[2, -1] ** f[2, 1] +
  2 i (t U) ** f†[2, -1] ** f†[2, 1] ** f[2, -1] ** f[3, 1] +
  2 i (t U) ** f†[2, -1] ** f†[2, 1] ** f[3, -1] ** f[2, 1] -
  2 i (t U) ** f†[2, -1] ** f†[3, 1] ** f[3, -1] ** f[3, 1] -
  2 i (t U) ** f†[3, -1] ** f†[2, 1] ** f[3, -1] ** f[3, 1] +
  2 i (t U) ** f†[3, -1] ** f†[3, 1] ** f[3, -1] ** f[4, 1] +
  2 i (t U) ** f†[3, -1] ** f†[3, 1] ** f[4, -1] ** f[3, 1] -
  2 i (t U) ** f†[3, -1] ** f†[4, 1] ** f[4, -1] ** f[4, 1] -
  2 i (t U) ** f†[4, -1] ** f†[1, 1] ** f[1, -1] ** f[4, 1] +
  2 i (t U) ** f†[4, -1] ** f†[3, 1] ** f[4, -1] ** f[4, 1] +
  2 i (t U) ** f†[4, -1] ** f†[4, 1] ** f[1, -1] ** f[4, 1] +
  2 i (t U) ** f†[4, -1] ** f†[4, 1] ** f[4, -1] ** f[1, 1]
```

Again, a system with an odd number of sites commutes with the Hamiltonian.

```
Timing[Commutator[H[7] /. PBCrule[7], jA[7] /. PBCrule[7]]]
```

{59.062, 0}

We now verify that $j_B$ commutes with the Hamiltonian for both odd and even number of sites.

```
Timing[Commutator[H[5] /. PBCrule[5], jB[5] /. PBCrule[5]]]
```

{42.806, 0}

```
Timing[Commutator[H[6] /. PBCrule[6], jB[6] /. PBCrule[6]]]
```

{62.323, 0}

Here we investigate whether the two currents commute with each other. The results suggest that for odd N the two currents commute but that they do not commute for even N.

```
Timing[Commutator[jA[5] /. PBCrule[5], jB[5] /. PBCrule[5]]]
```

{95.364, 0}

From the computation below we see that the commutator is non zero for even sites.

```
evensitecommutator = Commutator[jA[4] /. PBCrule[4], jB[4] /. PBCrule[4]];
```

```
Short[evensitecommutator, 2]
```

```
2 (t U) ** f†[1, -1] ** f†[1, 1] ** f[1, -1] ** f[3, 1] -
  2 ≪1≫ + ≪44≫ + 2 (t U) ** f†[4, -1] ** f†[4, 1] ** f[4, -1] ** f[2, 1]
```

In view of the above results, it seems very likely that $[j_A, T] = 0$ for odd number of sites, where T is the transfer matrix of the Hubbard model in the Fermi representation, whereas the commutator is non zero for even sites.

## ■ (VIII)  Construction of Cluster Hamiltonians of the Hubbard model using Bras and Kets

We illustrate the use of the Bra[Vacuum] and Ket[Vacuum] symbols, for creating the space of allowed states in the Hubbard model on a small cluster, and furhter to operate the Hamiltonian on these to generate its matrix representation. Let us consider a 4 site cluster defined in the diagram, with the Hubbard Hamiltonian hopping parameters (t1, t2 , U). For illustration we confine ourselves to the sector with Sz=0 and 4 particles, i.e. the half filled state, where the number of basis states is 36.

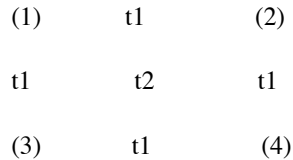

4-Site Hubbard Cluster with nearest neighbour hops t1, and second neighbour hops t2.

In[3]:= **hop[i_, j_] := - Sum[f†[i, σ] ** f[j, σ] + f†[j, σ] ** f[i, σ] , {σ, -1, 1, 2}]**

In[4]:= **pot[i_] := U n_f[i, 1] ** n_f[i, -1]**

In[5]:= **Hcluster = t1 ( hop[1, 2] + hop[2, 3] + hop[3, 4] + hop[4, 1]) +**
        **t2 ( hop[1, 4] + hop[2, 3])  + pot[1] + pot[2] + pot[3] + pot[4] ;**

We now define the creation and destruction operators for pairs of sites with a given spin projection,

In[6]:= **twofermi[i_, j_, σ_] = f†[i, σ] ** f†[j, σ];**
      **twofermidest[i_, j_, σ_] = f[j, σ] ** f[i, σ];**

Let us see the explicit form of a basis ket. The function StandardOrderQ organizes a Fermi operator product, its convention is to write these with increasing site labels from left to right, and puts the down spin blocks to the left of the up spin blocks.

In[8]:= **StandardOrderQ[twofermi[1, 2, 1] ** twofermi[2, 3, -1] ** Ket[Vacuum]]**

Out[8]= f†[2, -1] ** f†[3, -1] ** f†[1, 1] ** f†[2, 1] ** |Vacuum⟩

We next produce the 36 basis kets {Ψ[1], .., Ψ[36]}    and their adjoint 36 basis bras {ΨB[1], .., ΨB[36]}

In[9]:= **ii = 1; Do[Φ[ii] = StandardOrderQ[twofermi[i, j, 1] ** twofermi[k, 1, -1] ** Ket[Vacuum]];**
      **ii = ii + 1, {i, 1, 4}, {j, 1, i-1}, {k, 1, 4}, {1, 1, k-1}];**
      **ii = 1; Do[ΦB[ii] = StandardOrderQ[**
        **Bra[Vacuum] ** twofermidest[i, j, 1] ** twofermidest[k, 1, -1]];**
      **ii = ii + 1, {i, 1, 4}, {j, 1, i-1}, {k, 1, 4}, {1, 1, k-1}];**

It is useful to inspect a typical basis state pair

In[11]:= **Φ[9]**

Out[11]= f†[2, -1] ** f†[3, -1] ** f†[1, 1] ** f†[3, 1] ** |Vacuum⟩

In[12]:= **ΦB[18]**

Out[12]= ⟨Vacuum| ** f[3, -1] ** f[4, -1] ** f[2, 1] ** f[3, 1]

We next illustrate the action of the Hubbard Hamiltonian on these states, here the function StandardOrderQ does the job of simplifying expressions bu pushing the destruction operators to the extreme right and hitting the Ket[Vacuum], which is annihilated, or a similar story for the Bra[Vacuum] that is annihilated by creation operators. A brief glance at the output shows that we get back the basis states multiplied by coefficients that depend on t1,t2 and U.

In[13]:= **Short[StandardOrderQ[Hcluster ** Ψ[9]], 3]**

Out[13]//Short= $-t1 ** f†[1, -1] ** f†[3, -1] ** f†[1, 1] ** f†[3, 1] ** |$Vacuum$\rangle -$
$t1 ** f†[2, -1] ** f†[3, -1] ** f†[1, 1] ** f†[2, 1] ** |$Vacuum$\rangle - \ll1\gg - \ll1\gg + \ll1\gg -$
$\ll1\gg - \ll1\gg + \ll1\gg + U ** f†[2, -1] ** f†[3, -1] ** f†[1, 1] ** f†[3, 1] ** |$Vacuum$\rangle$

In[14]:= **Short[StandardOrderQ[ΨB[18] ** Hcluster], 2]**

Out[14]//Short= $t1 ** \langle$Vacuum$| ** f[1, -1] ** f[3, -1] ** f[2, 1] ** f[3, 1] - \ll1\gg - \ll1\gg -$
$\ll1\gg + \ll1\gg - \ll1\gg + U ** \langle$Vacuum$| ** f[3, -1] ** f[4, -1] ** f[2, 1] ** f[3, 1]$

At this point, there are two ways to construct the Hamiltonian matrix. Starting with expressions of the type StandardOrderQ[Hcluster**Ψ[i]], we can take the inner product with the bras ΨB[j]. As an alternative, we can pick off the coefficients using the QCoefficient function, this is faster and hence prefereable in most situatiions. We illustrate both methods and display the time saved by the second method. First we need a rule that collapses the fundamental inner-product to unity:

In[15]:= **ruleinnerproduct = {Bra[Vacuum] ** Ket[Vacuum] → 1};**

In[16]:= **Timing[Do[res[i] = StandardOrderQ[Hcluster ** Ψ[i]], {i, 1, 36}]]**

Out[16]= {4.5608, Null}

The array res[i] stores the resultant of the action of the Hamiltonian on the ith ket.
The two methods to creat the Hamiltonian matrix use the innerproduct with the ΨB[j], i.e. the Bra vector, or picking out the coefficients of Ψ[j] in res[i]. These are called matrix1 and matrix2 respectively.

**matrix1[j_, i_] := StandardOrderQ[ΨB[j] ** res[i]] /. ruleinnerproduct**
**matrix2[j_, i_] := QCoefficient[res[i], Ψ[j]]**

In[34]:= **R1 = Timing[Table[matrix1[k, i], {k, 1, 36}, {i, 1, 36}]];**
**R2 = Timing[Table[matrix2[k, i], {k, 1, 36}, {i, 1, 36}]];**

In[36]:= **R1[[1]]**

Out[36]= 401.36

In[37]:= **R2[[1]]**

Out[37]= 7.00216

In[39]:= **Short[R1[[2]] - R2[[2]], 3]**

Out[39]//Short= {{0, 0, 0, 0, 0, 0, 0, 0, 0, 0, 0, 0, 0, 0, 0, 0, 0, 0, 0, 0, 0, 0, 0, 0,
0, 0, 0, 0, 0, 0, 0, 0, 0, 0, 0, 0}, ≪34≫, {0, 0, 0, 0, 0, 0, 0, 0, 0, 0,
0, 0, 0, 0, 0, 0, 0, 0, 0, 0, 0, 0, 0, 0, 0, 0, 0, 0, 0, 0, 0, 0, 0, 0, 0, 0}}

We see that the two methods give the same resulting matrix, but the second method is considerably faster. We next define the ClusterHamiltonian and check its eigenvalues in some simple cases. For larger systems, it is clearly more advantageous to use the sparse matrix notation in such problems, but we will not pursue it here.

In[40]:= **ClusterHamiltonian[t1_, t2_, U_] = R1[[2]];**

In[41]:= **Eigenvalues[ClusterHamiltonian[1, 0, 0]]**

Out[41]= {-4, -4, -4, -4, 4, 4, 4, 4, -2, -2, -2, -2, -2, -2,
-2, -2, 2, 2, 2, 2, 2, 2, 2, 2, 0, 0, 0, 0, 0, 0, 0, 0, 0, 0, 0, 0}

In[42]:= **Eigenvalues[ClusterHamiltonian[0, 0, U]]**

Out[42]= {0, 0, 0, 0, 0, 0, U, U, U, U, U, U, U, U, U, U,
U, U, U, U, U, U, U, U, U, U, U, U, U, U, 2 U, 2 U, 2 U, 2 U, 2 U, 2 U}

The first set of eigenvalues correspond to a non interacting model with only nearest neighbour hoppings, and is easily seen to be correct. The second set of eigenvalues correspond to the local limit, where we have 6 states in the U=0

sector (Lowest Hubbard Band), 6 states in the U=2 sector (Second Hubbard band) and the rest with U=1, i.e. the First Hubbard Band.

---

## Appendices

### ■ (I) Explanation of Form

The strength of the system lies in the ability to organize, rearrange, and manipulate terms in an expression based on their properties. The method of organization provides a simple language for the manipulation of terms. An understanding of the language used by the code is not essential for most operations. The user is only required to input algebraic expressions in a logical and minimally restricted form to effectively use the functions of this package. The language of organization will briefly be covered here for posterity.

*Mathematica* organizes all input in terms of lists. Each list has a head. The head of a list is the function that is to be applied to the list. The example below shows how *Mathematica* organizes some algebraic input into lists and functions.

$$\text{FullForm}\left[5 \sum_i c[i]\, b[i] ** b^\dagger\,[j] + 2 \sum_i d[i]\, b^\dagger\,[i] ** b[j]\right]$$

Plus[Times[5, Sum[Times[c[i], NonCommutativeMultiply[b[i], b\[Dagger][j]]], i]],
  Times[2, Sum[Times[d[i], NonCommutativeMultiply[b\[Dagger][i], b[j]]], i]]]

It is the job of the Organize function to scan through this list and extract the relevant information. The Organize function is therefore at the heart of every operation of the code. This function reads input such as the example above and organizes it based on the properties of the terms encountered. Information encountered by the Organize function is stored in nested lists. Different types of objects are placed in different places within the larger list created by the Organize function. Organize recognizes four types of objects : numbers, sums, constants, and operators. Each of these types of objects is placed in a separate nested list. The easiest way to understand this ordering is through example

$$\text{Organize}\left[5 \sum_i c[i]\, b[i] ** b^\dagger\,[j]\right]$$

{{5, {i}, {c[i]}, {b[i], b†[j]}}}



Fig. (1): The output of organize is a list of lists. Each list corresponds to a specific type of term.

**Organize**$\left[5 \sum_i c[i]\, b[i] ** b^\dagger[j] + 2 \sum_i d[i]\, b^\dagger[i] ** b[j]\right]$

{{5, {i}, {c[i]}, {b[i], b†[j]}}, {2, {i}, {d[i]}, {b†[i], b[j]}}}

Fig.(2) : When the input of Organize is the addition of several terms,

each term is contained in a list within the larger list.

When a function is encountered that cannot be decomposed further, the Organize function determines whether this function contains any operators. If so, the package uses a special notation to signify that it has found a function of operators that cannot be decomposed further. The notation function[a, {b}] is used, where a is the function that cannot be decomposed, and b is a list of the operators on which the function depends. As it currently stands not all functions can be read and placed in this notation, but any function involving operators to different powers or exponential functions of operators can be understood. Using this format the user is free to define commutators of more complicated functions of operators as required. The notation used is identical to that used to add an operator, with the caveat that the definition will be written using the "function" notation described above. An example of an organized function such as described above is given below.

`Organize[a e^{t[i] q[i]+q[j]}]`

`{{1, {}, {a}, {function[e^{q[j]}, {q[j]}], function[e^{q[i] t[i]}, {q[i]}]}}}`

Humanize is the functional opposite of organize in that Humanize reads the language created by the Organize function and recreates a mathematical form that a user can understand. This function is demonstrated below.

`Humanize[{{5, {i}, {c[i]}, {b[i], b†[j]}}, {2, {i}, {d[i]}, {b†[i], b[j]}}}]`

$$5 \sum_i c[i] ** b[i] ** b†[j] + 2 \sum_i d[i] ** b†[i] ** b[j]$$

With this simple language the manipulation of certain types of terms can be performed with greater ease. Rather than manipulating operators as we come across them, we collect every operator in the expression. We can then perform algorithms on the lists of operators to combine and manipulate them as necessary. It is necessary to collect numbers and constants separately so that numbers can be combined as necessary and constants, which may depend on indices of summation, can be placed inside a sum. Also, the ability to identify summed indices allows us to evaluate delta functions. Every function in the package uses this form to organize input.

- **(II) Demonstration Problem Output (Produces large outputs hence not meant for printing)**


\end{document}